\documentclass[10pt,a4paper,onecolumn]{article}
\usepackage{subcaption}
\usepackage{marginnote}
\usepackage{graphicx}
\usepackage{xcolor}
\usepackage{authblk,etoolbox}
\usepackage{titlesec}
\usepackage{calc}
\usepackage{tikz}
\usepackage{hyperref}
\hypersetup{colorlinks,breaklinks=true,
            urlcolor=[rgb]{0.0, 0.5, 1.0},
            linkcolor=[rgb]{0.0, 0.5, 1.0}}
\usepackage{caption}
\usepackage{tcolorbox}
\usepackage{amssymb,amsmath}
\usepackage{ifxetex,ifluatex}
\usepackage{seqsplit}
\usepackage{xstring}

\usepackage{float}
\let\origfigure\figure
\let\endorigfigure\endfigure
\renewenvironment{figure}[1][2] {
    \expandafter\origfigure\expandafter[H]
} {
    \endorigfigure
}

\usepackage{fixltx2e} 
\usepackage[
  backend=biber,
]{biblatex}
\bibliography{paper.bib}

\let\textttOrig=\texttt
\def\texttt#1{\expandafter\textttOrig{\seqsplit{#1}}}
\renewcommand{\seqinsert}{\ifmmode
  \allowbreak
  \else\penalty6000\hspace{0pt plus 0.02em}\fi}

\makeatletter
\let\href@Orig=\href
\def\href@Urllike#1#2{\href@Orig{#1}{\begingroup
    \def\Url@String{#2}\Url@FormatString
    \endgroup}}
\def\href@Notdoi#1#2{\def\tempa{#1}\def\tempb{#2}%
  \ifx\tempa\tempb\relax\href@Urllike{#1}{#2}\else
  \href@Orig{#1}{#2}\fi}
\def\href#1#2{%
  \IfBeginWith{#1}{https://doi.org}%
  {\href@Urllike{#1}{#2}}{\href@Notdoi{#1}{#2}}}
\makeatother

\usepackage[top=3.5cm, bottom=3cm, right=1.5cm, left=1.0cm,
            headheight=2.2cm, reversemp, includemp, marginparwidth=4.5cm]{geometry}

\titleformat{\section}
  {\normalfont\sffamily\Large\bfseries}
  {}{0pt}{}
\titleformat{\subsection}
  {\normalfont\sffamily\large\bfseries}
  {}{0pt}{}
\titleformat{\subsubsection}
  {\normalfont\sffamily\bfseries}
  {}{0pt}{}
\titleformat*{\paragraph}
  {\sffamily\normalsize}

\usepackage{fancyhdr}
\makeatletter
\let\ps@plain\ps@fancy
\fancyheadoffset[L]{4.5cm}
\fancyfootoffset[L]{4.5cm}

\definecolor{linky}{rgb}{0.0, 0.5, 1.0}

\newtcolorbox{repobox}
   {colback=red, colframe=red!75!black,
     boxrule=0.5pt, arc=2pt, left=6pt, right=6pt, top=3pt, bottom=3pt}

\newcommand{\ExternalLink}{%
   \tikz[x=1.2ex, y=1.2ex, baseline=-0.05ex]{%
       \begin{scope}[x=1ex, y=1ex]
           \clip (-0.1,-0.1)
               --++ (-0, 1.2)
               --++ (0.6, 0)
               --++ (0, -0.6)
               --++ (0.6, 0)
               --++ (0, -1);
           \path[draw,
               line width = 0.5,
               rounded corners=0.5]
               (0,0) rectangle (1,1);
       \end{scope}
       \path[draw, line width = 0.5] (0.5, 0.5)
           -- (1, 1);
       \path[draw, line width = 0.5] (0.6, 1)
           -- (1, 1) -- (1, 0.6);
       }
   }

\patchcmd{\@maketitle}{center}{flushleft}{}{}
\patchcmd{\@maketitle}{center}{flushleft}{}{}
\patchcmd{\@maketitle}{\LARGE}{\LARGE\sffamily}{}{}
\def\maketitle{{%
  
  \AB@maketitle}}
\makeatletter
\renewcommand\AB@affilsepx{ \protect\Affilfont}
\renewcommand\AB@affilnote[1]{{\bfseries #1}\hspace{3pt}}
\renewcommand{\affil}[2][]%
   {\newaffiltrue\let\AB@blk@and\AB@pand
      \if\relax#1\relax\def\AB@note{\AB@thenote}\else\def\AB@note{#1}%
        \setcounter{Maxaffil}{0}\fi
        \begingroup
        \let\href=\href@Orig
        \let\texttt=\textttOrig
        \let\protect\@unexpandable@protect
        \def\thanks{\protect\thanks}\def\footnote{\protect\footnote}%
        \@temptokena=\expandafter{\AB@authors}%
        {\def\\{\protect\\\protect\Affilfont}\xdef\AB@temp{#2}}%
         \xdef\AB@authors{\the\@temptokena\AB@las\AB@au@str
         \protect\\[\affilsep]\protect\Affilfont\AB@temp}%
         \gdef\AB@las{}\gdef\AB@au@str{}%
        {\def\\{, \ignorespaces}\xdef\AB@temp{#2}}%
        \@temptokena=\expandafter{\AB@affillist}%
        \xdef\AB@affillist{\the\@temptokena \AB@affilsep
          \AB@affilnote{\AB@note}\protect\Affilfont\AB@temp}%
      \endgroup
       \let\AB@affilsep\AB@affilsepx
}
\makeatother

\renewcommand\Affilfont{\sffamily\small\mdseries}
\ifnum 0\ifxetex 1\fi\ifluatex 1\fi=0 
  \usepackage[T1]{fontenc}
  \usepackage[utf8]{inputenc}

\else 
  \ifxetex
    \usepackage{mathspec}
    \usepackage{fontspec}

  \else
    \usepackage{fontspec}
  \fi
  \defaultfontfeatures{Ligatures=TeX,Scale=MatchLowercase}

\fi
\IfFileExists{upquote.sty}{\usepackage{upquote}}{}
\IfFileExists{microtype.sty}{%
\usepackage{microtype}
\UseMicrotypeSet[protrusion]{basicmath} 
}{}

\usepackage{hyperref}
\hypersetup{unicode=true,
            pdftitle={Fast fully-reproducible streamlined serial/parallel Monte Carlo/MCMC simulations and visualizations via ParaMonte::Python library},
            pdfborder={0 0 0},
            breaklinks=true}
\usepackage{color}
\usepackage{fancyvrb}

\DefineVerbatimEnvironment{Highlighting}{Verbatim}{commandchars=\\\{\}}
\newenvironment{Shaded}{}{}
\newcommand{\KeywordTok}[1]{\textcolor[rgb]{0.00,0.44,0.13}{\textbf{#1}}}

\newcommand{\DecValTok}[1]{\textcolor[rgb]{0.25,0.63,0.44}{#1}}

\newcommand{\FloatTok}[1]{\textcolor[rgb]{0.25,0.63,0.44}{#1}}

\newcommand{\ImportTok}[1]{#1}
\newcommand{\CommentTok}[1]{\textcolor[rgb]{0.38,0.63,0.69}{\textit{#1}}}

\newcommand{\VariableTok}[1]{\textcolor[rgb]{0.10,0.09,0.49}{#1}}
\newcommand{\ControlFlowTok}[1]{\textcolor[rgb]{0.00,0.44,0.13}{\textbf{#1}}}
\newcommand{\OperatorTok}[1]{\textcolor[rgb]{0.40,0.40,0.40}{#1}}
\newcommand{\BuiltInTok}[1]{#1}
\newcommand{\ExtensionTok}[1]{#1}

\newcommand{\NormalTok}[1]{#1}

\let\addcontentslineOrig=\addcontentsline
\def\addcontentsline#1#2#3{\bgroup
  \let\texttt=\textttOrig\addcontentslineOrig{#1}{#2}{#3}\egroup}
\let\markbothOrig\markboth
\def\markboth#1#2{\bgroup
  \let\texttt=\textttOrig\markbothOrig{#1}{#2}\egroup}
\let\markrightOrig\markright
\def\markright#1{\bgroup
  \let\texttt=\textttOrig\markrightOrig{#1}\egroup}

\usepackage{graphicx,grffile}
\makeatletter
\def\maxwidth{\ifdim\Gin@nat@width>\linewidth\linewidth\else\Gin@nat@width\fi}
\def\maxheight{\ifdim\Gin@nat@height>\textheight\textheight\else\Gin@nat@height\fi}
\makeatother
\setkeys{Gin}{width=\maxwidth,height=\maxheight,keepaspectratio}
\IfFileExists{parskip.sty}{%
\usepackage{parskip}
}{
\setlength{\parindent}{0pt}
\setlength{\parskip}{6pt plus 2pt minus 1pt}
}
\providecommand{\tightlist}{%
  \setlength{\itemsep}{0pt}\setlength{\parskip}{0pt}}
\ifx\paragraph\undefined\else
\let\oldparagraph\paragraph
\renewcommand{\paragraph}[1]{\oldparagraph{#1}\mbox{}}
\fi
\ifx\subparagraph\undefined\else
\let\oldsubparagraph\subparagraph
\renewcommand{\subparagraph}[1]{\oldsubparagraph{#1}\mbox{}}
\fi

\title{Fast fully-reproducible streamlined serial/parallel Monte Carlo/MCMC
simulations and visualizations via \texttt{ParaMonte::Python} library}

        \author[1, 2]{Amir Shahmoradi}
          \author[1]{Fatemeh Bagheri}
          \author[1]{Joshua Alexander Osborne}
    
      \affil[1]{Department of Physics, The University of Texas, Arlington, TX}
      \affil[2]{Data Science Program, The University of Texas, Arlington, TX}
  \date{\vspace{-7ex}}

\begin{document}
\maketitle

\marginpar{

  \begin{flushleft}
  \sffamily\small

  {\bfseries DOI:} \href{https://doi.org/}{\color{linky}{}}

  \vspace{2mm}

  {\bfseries Software}
  \begin{itemize}
    \setlength\itemsep{0em}
    \item \href{}{\color{linky}{Review}} \ExternalLink
    \item \href{https://github.com/cdslaborg/paramonte/tree/master/src/interface/Python}{\color{linky}{Repository}} \ExternalLink
    \item \href{}{\color{linky}{Archive}} \ExternalLink
  \end{itemize}

  \vspace{2mm}

  \par\noindent\hrulefill\par

  \vspace{2mm}

  {\bfseries Editor:} \href{}{} \ExternalLink \\
  \vspace{1mm}
    \vspace{2mm}

  {\bfseries Submitted:} 29 September 2020\\
  {\bfseries Published:} 

  \vspace{2mm}
  {\bfseries License}\\
  Authors of papers retain copyright and release the work under a Creative Commons Attribution 4.0 International License (\href{http://creativecommons.org/licenses/by/4.0/}{\color{linky}{CC BY 4.0}}).

  \end{flushleft}
}

\section{Summary}\label{summary}

\texttt{ParaMonte::Python} (standing for \textbf{Para}llel
\textbf{Monte} Carlo in \textbf{Python}) is a serial and
MPI-parallelized library of (Markov Chain) Monte Carlo (MCMC) routines
for sampling mathematical objective functions, in particular, the
posterior distributions of parameters in Bayesian modeling and analysis
in data science, Machine Learning, and scientific inference. In addition
to providing access to fast high-performance serial/parallel Monte Carlo
and MCMC sampling routines, the \texttt{ParaMonte::Python} library
provides extensive post-processing and visualization tools that aim to
automate and streamline the process of model calibration and uncertainty
quantification in Bayesian data analysis. Furthermore, the
automatically-enabled restart functionality of
\texttt{ParaMonte::Python} samplers ensures seamless fully-deterministic
into-the-future restart of Monte Carlo simulations, should any runtime
interruptions happen. The \texttt{ParaMonte::Python} library is
MIT-licensed and is
\href{https://github.com/cdslaborg/paramonte/tree/master/src/interface/Python}{permanently
maintained on GitHub}.

\section{Statement of need}\label{statement-of-need}

Originally developed in 1949 (Metropolis \& Ulam, 1949), Monte Carlo
simulation techniques, in particular, the Markov Chain Monte Carlo
(MCMC) have become one of the most popular methods of uncertainty
quantification in quantitative research. A number of Python packages
already provide probabilistic programming environments for Markov Chain
Monte Carlo simulations within which the user is expected to implement
their inference problems in the specific language and syntax designed
for the package (Patil, Huard, \& Fonnesbeck, 2010), (Team \& others,
2017). While such approaches to Monte Carlo simulations can facilitate
the implementation of simple inference problems, they can potentially
limit the user's ability to implement sophisticated mathematical models
whose complexities go beyond what these probabilistic programming
language environments can offer.

Other MCMC packages require the objective function to be provided by the
user as a black-box with a pre-specified procedural interface
(Foreman-Mackey et al., 2019) or require custom definitions for the
objective function (Miles, 2019). Regardless of the interface
requirements, the majority of the existing Python MCMC environments are
limited to Python programming language. Exceptions include the PyStan
probabilistic programming environment and \texttt{pymcmcstat} which is
also available from C++.

The \texttt{ParaMonte::Python} library presented in this manuscript aims
to fill some of the gaps in the existing tools for (Markov Chain) Monte
Carlo simulations. We have built the \texttt{ParaMonte::Python} package
upon the ParaMonte kernel library which provides high-performance
serial/parallel MCMC simulation environment for the C, C++, and Fortran
programming languages. Along with \texttt{ParaMonte::MATLAB} and the
kernel C/C++/Fortran libraries, \texttt{ParaMonte::Python} aims to
provide a unified Application Programming Interface (API) to a range of
(Markov Chain) Monte Carlo samplers with a similar syntax and usage
across several programming languages. This is particularly true about
the MATLAB and Python languages, where the syntax of the ParaMonte
library as well as the visualization and post-processing tools look and
feel almost identical.

As the name of the library indicates, a particular focus of the
\texttt{ParaMonte::Python} library is to enable scalable parallel Monte
Carlo simulations on distributed as well as shared memory architectures.
In designing the ParaMonte library, we have followed the \emph{principle
of separation of concerns} to separate the computational implementation
of objective function from the implementation of the Monte Carlo
samplers. In addition, we have developed the library while bearing the
following principal design goals in mind:

\begin{itemize}
\tightlist
\item
  \textbf{Full automation} of the Monte Carlo simulations while
  providing extensive descriptions and guidance to the user, in
  real-time, to ensure the highest level of user-friendliness of the
  package for running, post-processing, and visualizing Monte Carlo and
  MCMC simulations.\\
\item
  \textbf{Unified-API} implementation of \texttt{ParaMonte::Python} to
  ensure the library looks and feels almost identical to the API of the
  ParaMonte library in other programming languages, for example,
  \texttt{ParaMonte::MATLAB}.\\
\item
  \textbf{High-Performance}, meticulously-low-level, implementation of
  the library's samplers, guaranteeing the fastest-possible Monte Carlo
  simulations, without compromising the reproducibility or the
  restart-functionality of the simulations.\\
\item
  \textbf{Parallelizability} of all simulations via distributed-memory
  MPI communications to ensure the scalability of simulations, from
  personal laptops to supercomputers, while \textbf{requiring
  zero-parallel-coding efforts from the user}.\\
\item
  \textbf{Zero external-library dependencies} of the kernel of the
  library, \textbf{other than \texttt{numpy}}, to ensure hassle-free
  library installation and Monte Carlo simulation runs. In practice,
  \texttt{ParaMonte::Python} has only one external-library dependency on
  \texttt{numpy} that is essential to perform Monte Carlo simulations,
  although post-processing and visualization of the results also require
  \texttt{scipy}, \texttt{pandas}, \texttt{matplotlib}, and
  \texttt{seaborn}. Some of the aforementioned Python libraries (e.g.,
  \texttt{seaborn} and \texttt{matplotlib}) tend to be unstable or not
  even available on some computing platforms, in particular, on
  supercomputers. Therefore, the installation of these libraries is
  intentionally \emph{not} enforced by the \texttt{ParaMonte::Python}
  installer script since these modules are only needed for the
  postprocessing and visualization of simulation results. Therefore, the
  user has the responsibility to install these libraries prior to using
  \texttt{ParaMonte::Python} for post-processing of simulation results.
  We have made sure to provide ample guidance and warnings to the user
  about this issue, when the \texttt{paramonte} module is imported to
  the user's Python environment for the first time.\\
\item
  \textbf{Fully-deterministic reproducibility} and
  \textbf{automatically-enabled restart functionality} for all Monte
  Carlo and MCMC simulations, which guarantee full recovery of
  simulations, should an interruption happen at any stage.\\
\item
  \textbf{Automatically-enabled comprehensive-reporting and
  post-processing} of the simulation results and their efficient compact
  storage in external files to ensure that simulation results will be
  reproducible and comprehensible in the future.
\end{itemize}

\section{Installation}\label{installation}

The \texttt{ParaMonte::Python} library is permanently maintained on
GitHub and is available at:
\url{https://github.com/cdslaborg/paramonte/tree/master/src/interface/Python}.
Each release of the library is also available on the The Python Package
Index (PyPI) repository. The most straightforward method of installation
is via \texttt{pip} on an Anaconda3 command-prompt on Windows or in a
Python-aware Bash terminal,

\begin{Shaded}
\begin{Highlighting}[]
\ExtensionTok{pip}\NormalTok{ install --user --upgrade paramonte}
\end{Highlighting}
\end{Shaded}

A primary goal in the design of the ParaMonte library has been the
automation of serial and parallel-scalable Monte Carlo simulations.
Therefore, upon the first \texttt{import\ paramonte} action in a Python
environment, the library checks for the existence of the MPI runtime
libraries on the user's system and if needed, automatically installs the
MPI library for parallel ParaMonte simulations. The entire process is
performed with the explicit permission from the user.

It is also possible to build the \texttt{ParaMonte::Python} library from
the source files. In such case, the entire build process of the library
is also fully automated. The one-line commands to automatically and
locally build the library on a given system are available on the
\href{https://www.cdslab.org/paramonte/notes/installation/python/}{documentation
website of the library}. The pre-built \texttt{ParaMonte::Python}
library also ships with a \texttt{build()} function that can
automatically build the library from source, locally, on the user's Unix
system. This includes the automatic installation of the GNU Compiler
collection, the MPI libraries, and any other build prerequisites (e.g.,
cmake, \ldots{}), if needed. All of these tasks are performed with the
explicit permission from the user.

\section{The ParaDRAM sampler}\label{the-paradram-sampler}

The current implementation of \texttt{ParaMonte::Python} library
contains the \textbf{\texttt{ParaDRAM}} sampler, standing for
\textbf{Para}llel \textbf{D}elayed-\textbf{R}ejection \textbf{A}daptive
\textbf{M}etropolis Markov Chain Monte Carlo (Shahmoradi \& Bagheri,
2020), (Shahmoradi \& Bagheri, 2020a), (Shahmoradi \& Bagheri, 2020b),
(Shahmoradi \& Bagheri, 2020c), (Kumbhare \& Shahmoradi, 2020).

The ParaDRAM sampler is a special variant of the DRAM algorithm by
(Haario, Laine, Mira, \& Saksman, 2006) and can be used in serial or
parallel mode. Unlike the traditional MCMC samplers where the proposal
distribution remains fixed throughout the simulation, the ParaDRAM
sampler ensures fast convergence to the objective function by
continuously adapting the proposal distribution of the MCMC sampler to
the shape of the objective function. The algorithm provides a highly
flexible and customizable simulation environment via an extensive number
of input specifications for the MCMC simulations. These simulation
specifications can be either set by the user, or left to be
appropriately determined by the sampler. A complete description of these
specifications go beyond the limits of this manuscript, but can be found
on
\href{https://www.cdslab.org/paramonte/notes/usage/paradram/specifications/}{the
documentation website of the library}. In addition, every ParaDRAM or
ParaMonte simulation automatically generates an output
\texttt{*\_report.txt} file containing the descriptions of all
simulation specifications as well as post-processing of the simulation
performance and results.

\subsection{Example Usage}\label{example-usage}

A major focus in the development of \texttt{ParaMonte::Python} has been
on the user-friendliness of the library and providing dynamic on-the-fly
user-guidance and directions toward the next step in the simulation
process. As such, setting up a ParaDRAM simulation, whether serially or
in parallel (via hundreds of supercomputer cores), and visualizing the
results can be achieved with minimal coding efforts by the user. For
example, sampling a Multivariate Normal distribution can be achieved via
the following code in serial mode,

\begin{Shaded}
\begin{Highlighting}[]
\ImportTok{import}\NormalTok{ numpy }\ImportTok{as}\NormalTok{ np}
\ImportTok{import}\NormalTok{ paramonte }\ImportTok{as}\NormalTok{ pm}
\KeywordTok{def}\NormalTok{ getLogFunc(point): }\ControlFlowTok{return} \OperatorTok{-}\FloatTok{0.5} \OperatorTok{*}\NormalTok{ np.}\BuiltInTok{sum}\NormalTok{( point}\OperatorTok{**}\DecValTok{2}\NormalTok{ )}
\NormalTok{pmpd }\OperatorTok{=}\NormalTok{ pm.ParaDRAM()}
\NormalTok{pmpd.runSampler ( ndim }\OperatorTok{=} \DecValTok{4} \CommentTok{# assume 4-dimensional objective function}
\NormalTok{                , getLogFunc }\OperatorTok{=}\NormalTok{ getLogFunc   }\CommentTok{# the objective function}
\NormalTok{                )}
\end{Highlighting}
\end{Shaded}

Each \texttt{ParaMonte::Python} simulation generates a number of output
files that contain information about one aspect of the simulation. A
complete description of these files is provided on
\href{https://www.cdslab.org/paramonte/notes/usage/paradram/output/}{the
documentation website of the library}. The library also provides a
number of visualization tools that aim to streamline the post-processing
of simulation results. For example, generating a \texttt{gridplot} of
all sampled states requires only one line of Python command,

\begin{figure}
\centering
\includegraphics{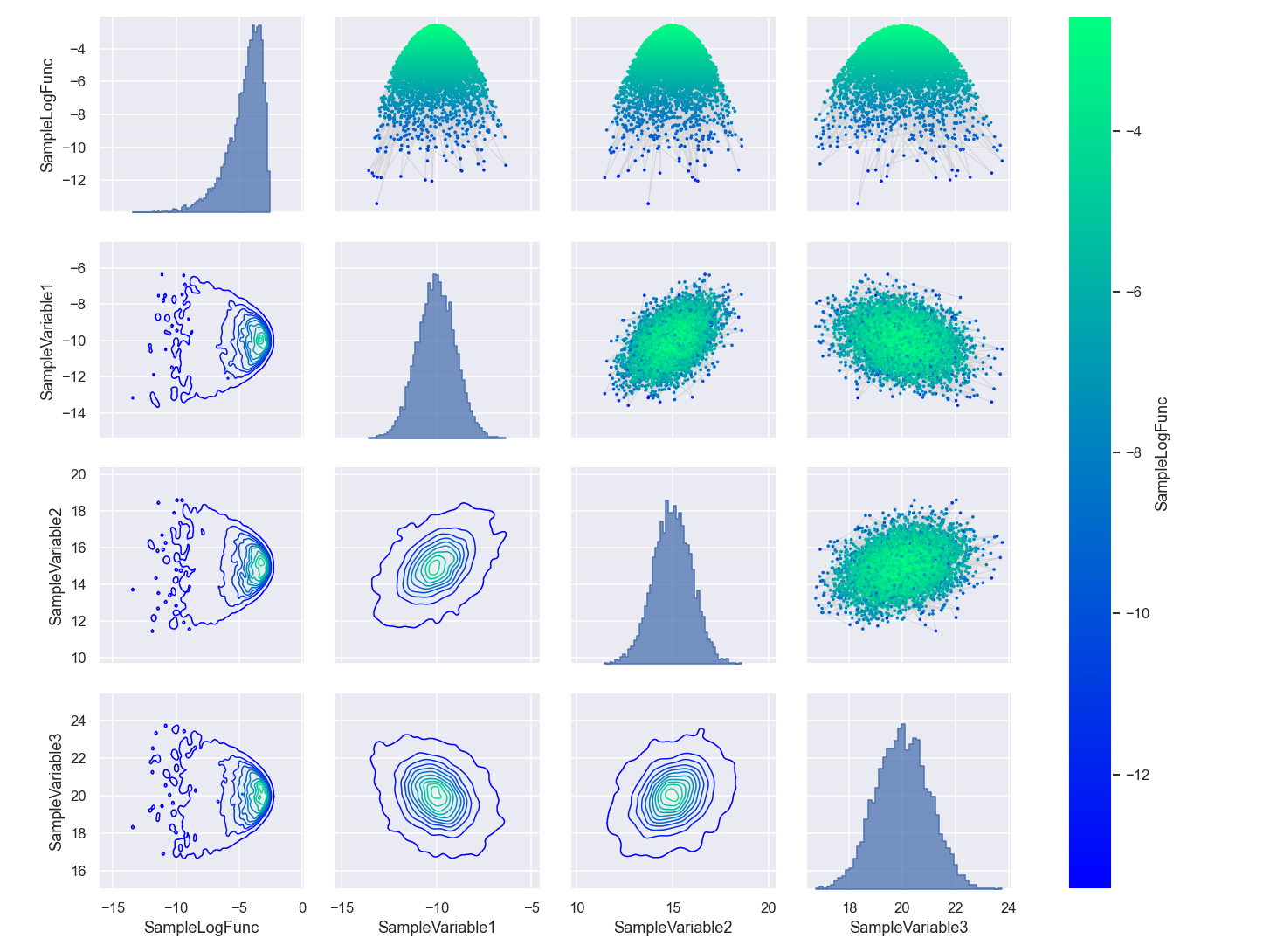}
\caption{An example \texttt{grid} plot displaying the results of
sampling a 4-dimensional MVN distribution with a random covariance
matrix.\label{fig:gridplot}}
\end{figure}

\begin{Shaded}
\begin{Highlighting}[]
\NormalTok{sample }\OperatorTok{=}\NormalTok{ pmpd.readSample(renabled}\OperatorTok{=}\VariableTok{True}\NormalTok{)[}\DecValTok{0}\NormalTok{] }\CommentTok{# input filename is optional}
\NormalTok{sample.plot.grid() }\CommentTok{# make a grid plot}
\end{Highlighting}
\end{Shaded}

An example output of the above command is shown in
\autoref{fig:gridplot} for sampling a 4-dimensional Multivariate Normal
(MVN) distribution. Each plotting tool of \texttt{ParaMonte::Python},
including the \texttt{grid} plot, contains a large number of attributes
that are automatically set to the appropriate default values to enable
quick streamlined visualizations. If desired, however, these default
values can be easily changed by the user to alter the behavior, display,
and contents of the resulting figures.

\subsection{Efficient compact storage of the
output}\label{efficient-compact-storage-of-the-output}

External storage of the output of Monte Carlo simulations is a highly
desired feature for both post-processing of the results and restarting
an interrupted simulation. The restart functionality is, in particular,
vital for large-scale computationally-expensive parallel Bayesian
inference problems. However, as the complexity of the target density and
the problem size increase, external storage of the simulation output can
quickly become a bottleneck in the computational speed of the
simulation.

The ParaDRAM sampler automatically generates multiple output files to
ensure seamless postprocessing and reproducibility of the results, in
particular, the automatic restart of interrupted simulations. To
minimize the effects of external input/output (IO) on the runtime speed
of the ParaDRAM sampler, we have implemented a novel method of carefully
storing the resulting MCMC chains from the sampler in a small,
\emph{compact}, yet human-readable (ASCII) format in external output
files.

The resulting output \textbf{compact-chain} (versus \textbf{verbose
Markov-chain}) format can lead to a significant speedup of the
simulation while demanding 4-100 times less external storage for the
simulation output. Additionally, the format of the output chain and
restart files can be also set to \texttt{binary}, further reducing the
memory foot-print of the simulation and increasing the simulation speed.
The implementation details of this compact-chain format go beyond the
scope of this paper, but are provided in (Shahmoradi \& Bagheri, 2020),
(Shahmoradi \& Bagheri, 2020b).

\subsection{Monitoring Convergence}\label{monitoring-convergence}

The continuous adaptation of the proposal distribution of the adaptive
Markov Chain Monte Carlo samplers is an issue that requires special
attention and care. Such adaptation, although increases the sampling
efficiency, can potentially break the reversibility and ergodic
properties of the Markov Chain. Theoretical results, however, indicate
that the convergence of the adaptive Markov chain to the target density
is guaranteed, as long as the adaptation of the Markov chain
monotonically decreases throughout the simulation (Haario et al., 2006).

It is, therefore, crucial to measure and dynamically monitor the amount
of adaptation in ParaDRAM simulations to ensure the adaptation of the
proposal distribution diminishes progressively throughout the
simulation. This is, however, a challenging task that requires the
quantification of the difference between two
potentially-multidimensional probability distributions. In (Shahmoradi
\& Bagheri, 2020), (Shahmoradi \& Bagheri, 2020b), we have introduced a
novel technique to circumvent this computationally NP-hard problem by
computing an upper bound on the total variation distance between any
subsequent pairs of adaptively-updated proposal distributions.

This \texttt{AdaptationMeasure} is a real number between 0 and 1 with 0
implying no proposal adaptation and 1 implying extreme adaptation. This
quantity is automatically written to the output \texttt{*\_chain.txt}
file for each ParaDRAM simulation, which can be explored and visualized
to ensure that the ergodicity and the reversibility properties of the
Markov chain hold.

\begin{figure}
\centering
\includegraphics{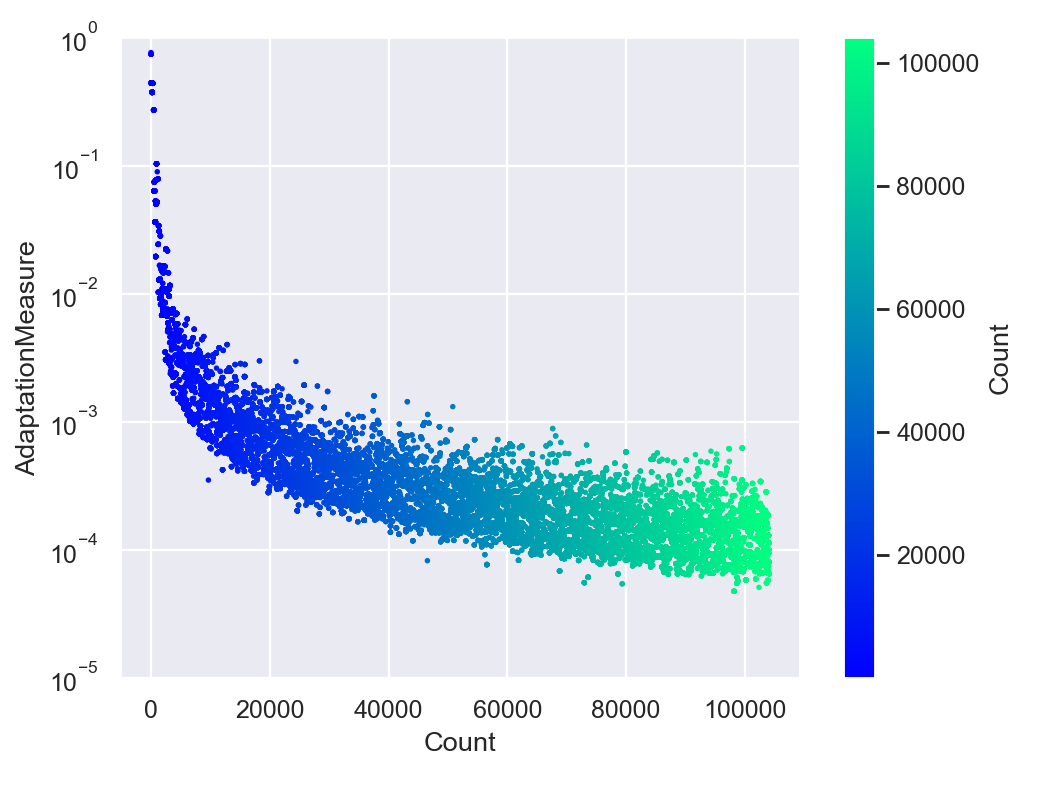}
\caption{An illustration of the diminishing-adaptation criterion of the
Delayed-Rejection Adaptive Metropolis Markov Chain Monte Carlo
(ParaDRAM) sampler of \texttt{ParaMonte::Python} library for the problem
of sampling a 4-dimensional Gaussian density
function.\label{fig:adaptMeasure}}
\end{figure}

\autoref{fig:adaptMeasure} illustrates the dynamics of
\texttt{AdaptationMeasure} for the problem of sampling a 4-dimensional
MVN density function. The observed behavior of
\texttt{AdaptationMeasure} in this figure is precisely the kind of
diminishing adaptation one would hope to witness in an adaptive Markov
Chain Monte Carlo simulation. Furthermore, a non-diminishing
\texttt{AdaptationMeasure} can be a strong indicator of lack of
convergence of the Markov chain to the target density function.

\begin{figure}
\centering
\includegraphics{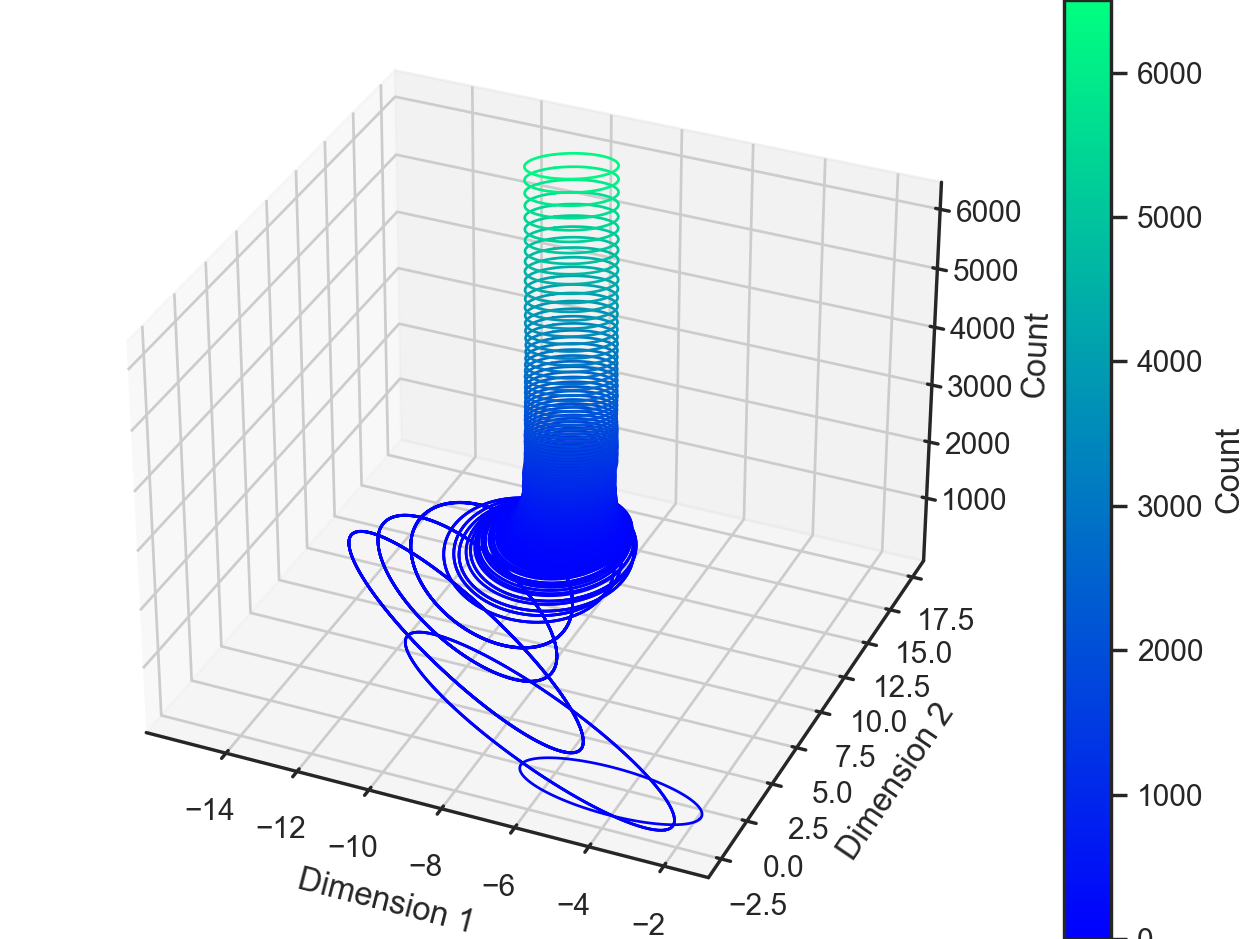}
\caption{A 3D illustration of the dynamic adaptation of the covariance
matrix of the 4-dimensional Gaussian proposal distribution of the
ParaDRAM sampler for an example problem of sampling a 4-dimensional MVN
target density function. As seen, the amount of adaptation of the
proposal distribution quickly converges to zero, ensuring the ergodicity
of the resulting Markov chain.\label{fig:covMatEvol}}
\end{figure}

\autoref{fig:covMatEvol} illustrates the dynamics of the adaptation of
the proposal distribution's covariance matrices throughout a ParaDRAM
sampling of the same 4-dimensional MVN target density function. The
information displayed in this figure is automatically saved in the
output \texttt{*\_restart.txt} file of every ParaDRAM simulation and can
be parsed via a single ParaDRAM command in Python,

\begin{Shaded}
\begin{Highlighting}[]
\NormalTok{restart }\OperatorTok{=}\NormalTok{ pmpd.readRestart(renabled}\OperatorTok{=}\VariableTok{True}\NormalTok{)[}\DecValTok{0}\NormalTok{]}
\end{Highlighting}
\end{Shaded}

The visualization of this information as depicted in
\autoref{fig:covMatEvol} can be achieved via another single-line Python
command,

\begin{Shaded}
\begin{Highlighting}[]
\NormalTok{restart.plot.covmat3()}
\end{Highlighting}
\end{Shaded}

Just as with other visualization tools of \texttt{ParaMonte::Python}
library, the aesthetics and contents of the plot can be readily
controlled via the attributes of the generated \emph{callable} Python
figure object \texttt{restart.plot.covmat3}.

\subsection{Sample refinement}\label{sample-refinement}

\includegraphics[width=0.49000\textwidth]{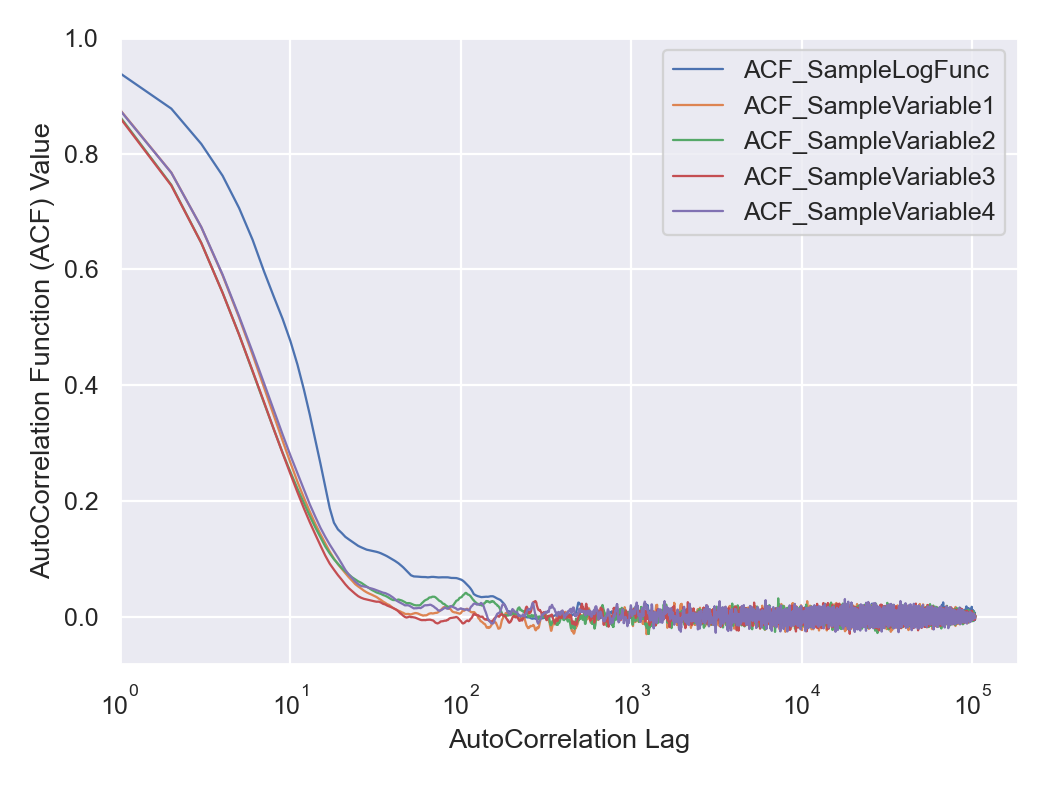}
\includegraphics[width=0.49000\textwidth]{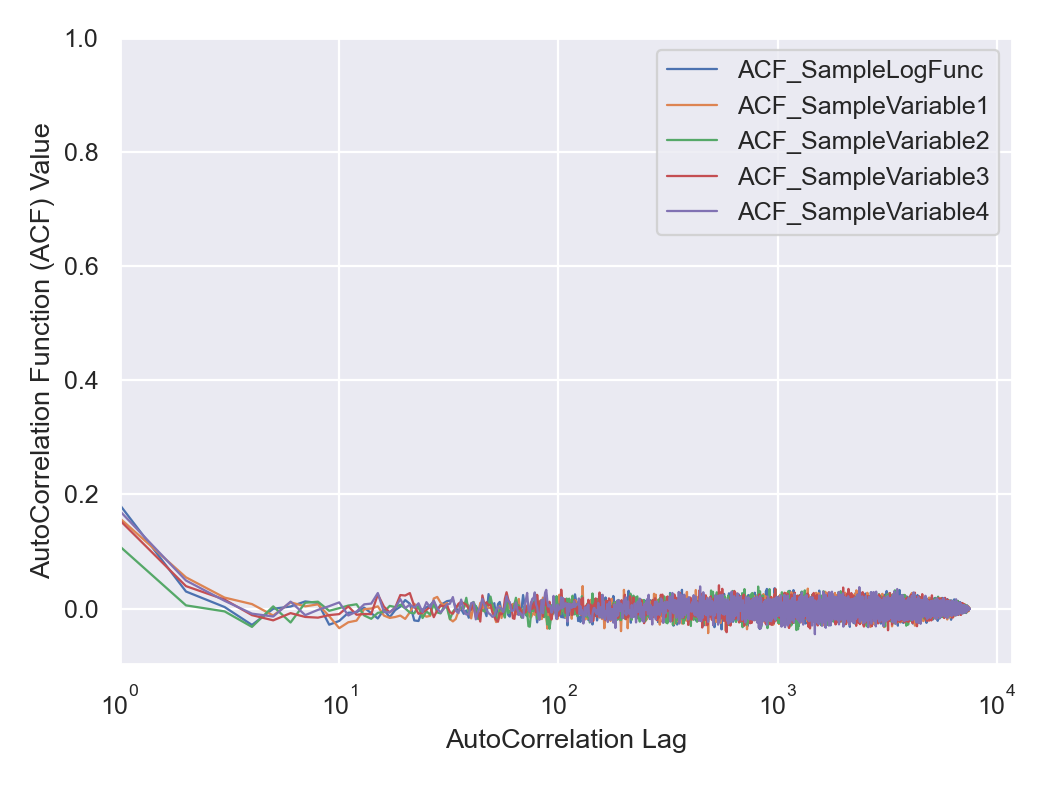}

\begin{figure}[!h]
    \begin{subfigure}[t]{0.49\textwidth}
        \caption{The Markov chain autocorrelation.}
    \end{subfigure}
    \hfill
    \begin{subfigure}[t]{0.49\textwidth}
        \caption{The refined-sample autocorrelation.}
    \end{subfigure}
    \caption{An illustration of the result of the aggressive and recursive refinements of the output Markov chain that is automatically performed by the ParaDRAM algorithm as part of post-processing of the results.}\label{fig:acf}
\end{figure}

The ParaDRAM algorithm automatically refines, aggressively and
recursively, each output Markov chain from a ParaDRAM simulation, such
that the resulting chain exhibits no autocorrelation. \autoref{fig:acf}
compares the autocorrelation of the raw verbose (Markov) chain from an
example ParaDRAM simulation run with the autocorrelation of the
corresponding refined sample. The recursive refinement procedure that we
have implemented in ParaDRAM is extensively detailed in (Shahmoradi \&
Bagheri, 2020), (Shahmoradi \& Bagheri, 2020a), (Shahmoradi \& Bagheri,
2020b).

\subsection{Parallelism}\label{parallelism}

The \texttt{ParaMonte::Python} parallel simulations are currently
enabled via the Message-Passing-Interface communication paradigm to
ensure the high-performance and scalability of the simulations across
multiple clusters of processors. The library can automatically detect
the presence of MPI runtime libraries required for parallel simulations
on the user's system. If any component is missing, it can also
automatically install the missing libraries with the explicit permission
from the user. The \texttt{ParaMonte::Python} samplers currently support
two modes of parallelism (Shahmoradi \& Bagheri, 2020), (Shahmoradi \&
Bagheri, 2020b),

\begin{itemize}
\item
  The \textbf{Perfect Parallelism} (multi-Chain), in which independent
  instances of the Markov Chain are simulated. Upon finishing the
  simulation, the ParaDRAM algorithm compares the output refined samples
  from all processors with each other to ensure that there is no
  evidence for a lack-of-convergence to the objective function.
\item
  The \textbf{Fork-Join Parallelism} (single-Chain), in which a single
  Markov chain is generated by the main processor, while other processes
  assist in proposing new states. This mode is the default behavior of
  the ParaDRAM sampler when used in parallel.
\end{itemize}

An example of a simple parallel ParaDRAM simulation sampling a
4-dimensional MVN is the following,

\begin{Shaded}
\begin{Highlighting}[]
\ImportTok{import}\NormalTok{ numpy }\ImportTok{as}\NormalTok{ np}
\ImportTok{import}\NormalTok{ paramonte }\ImportTok{as}\NormalTok{ pm}
\KeywordTok{def}\NormalTok{ getLogFunc(point): }\ControlFlowTok{return} \OperatorTok{-}\FloatTok{0.5} \OperatorTok{*}\NormalTok{ np.}\BuiltInTok{sum}\NormalTok{( point}\OperatorTok{**}\DecValTok{2}\NormalTok{ )}
\NormalTok{pmpd }\OperatorTok{=}\NormalTok{ pm.ParaDRAM()}
\NormalTok{pmpd.mpiEnabled }\OperatorTok{=} \VariableTok{True}
\NormalTok{pmpd.runSampler ( ndim }\OperatorTok{=} \DecValTok{4} \CommentTok{# assume 4-dimensional objective function}
\NormalTok{                , getLogFunc }\OperatorTok{=}\NormalTok{ getLogFunc   }\CommentTok{# the objective function}
\NormalTok{                )}
\end{Highlighting}
\end{Shaded}

Compared with serial simulations, the only extra piece of information
that is required from the user is the Python statement
\texttt{pmpd.mpiEnabled\ =\ True} in the above script. Other than
setting this variable, the \texttt{ParaMonte::Python} samplers require
\emph{zero parallel coding efforts from the user} to run simulations in
parallel.

Assuming the above Python script is saved in a file named
\texttt{main.py}, running the simulation in parallel on 3 processes is a
one-line command in a bash or Anaconda3 terminal,

\begin{Shaded}
\begin{Highlighting}[]
\ExtensionTok{mpiexec}\NormalTok{ -n 3 python main.py}
\end{Highlighting}
\end{Shaded}

Depending on the platform (in particular, on Windows and
supercomputers), the calling syntax of the MPI runtime library might be
slightly different. Comprehensive details and example simulations in
parallel on a variety of platforms are provided in
\href{https://www.cdslab.org/paramonte/}{the documentation of the
library}.

\subsubsection{Setting the optimal number of
processors}\label{setting-the-optimal-number-of-processors}

The ParaMonte samplers automatically compute and output the parallel
speedup compared to the serial mode. In addition, the samplers
automatically predict the optimal number of processors for the
simulation in parallel Fork-Join mode, given the characteristics of the
current parallel simulation run. These computations and predictions are
automatically stored in the \texttt{*\_report.txt} files that accompany
each simulation.

Although the predicted optimal number of processors is a first-order
approximation, it can be a very useful guide to set the number of
processes for large-scale computationally demanding simulations. In such
cases, the user can run a short parallel simulation with an arbitrary
number of processors and check the output report file of the simulation
to gain insight into the optimal number of processes for the
full-production parallel simulation. A complete description of the
algorithm that approximates the optimal number of processes is given in
(Shahmoradi \& Bagheri, 2020), (Shahmoradi \& Bagheri, 2020b).

\begin{figure}
\centering
\includegraphics{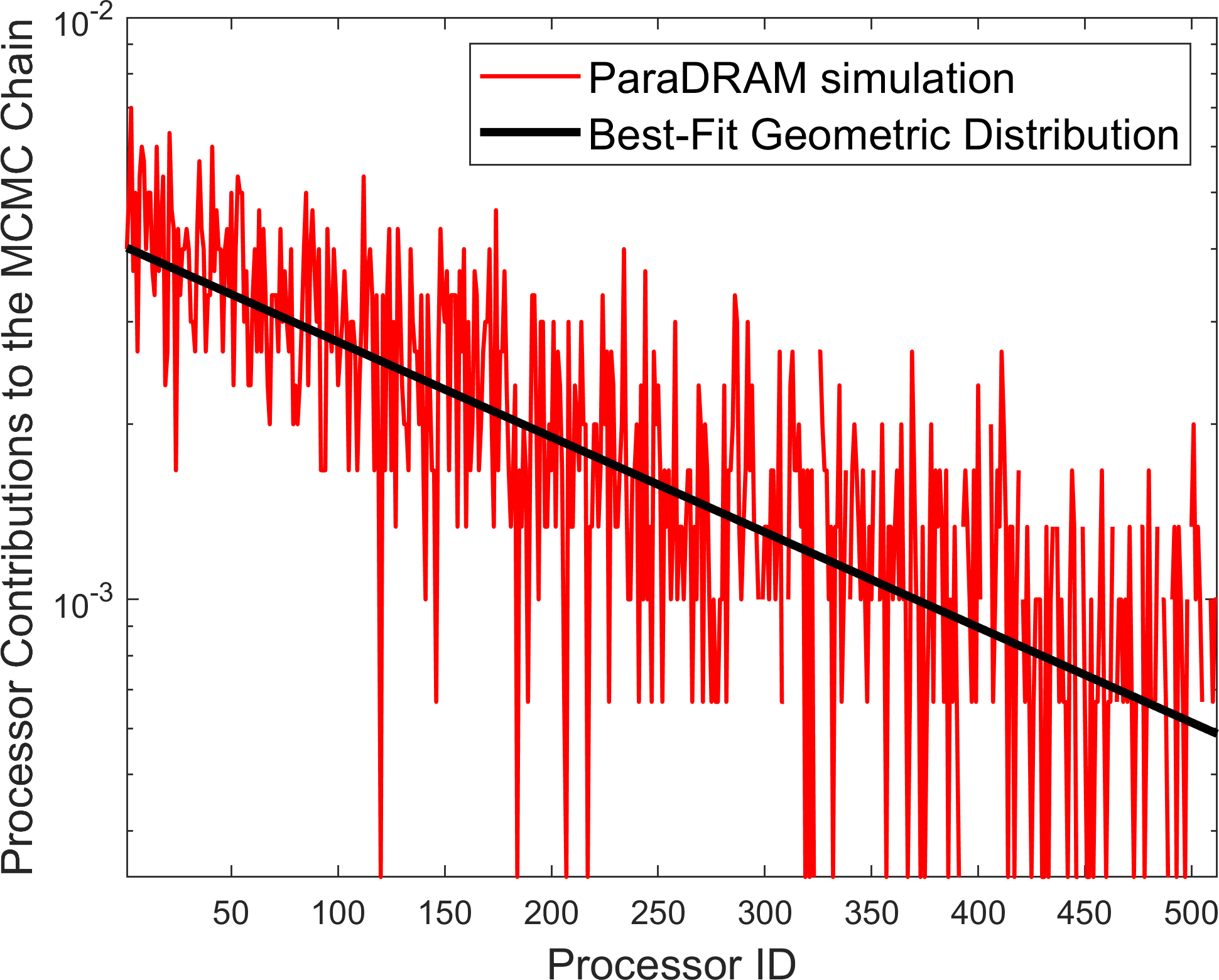}
\caption{The distribution of the contributions of 512 processors to a
ParaMonte::Python ParaDRAM simulation via the Fork-Join paradigm. The
balck line represents the best-fit Geometric distribution predicted in
the post-processing phase of the ParaDRAM simulation (Shahmoradi \&
Bagheri, 2020), (Shahmoradi \& Bagheri, 2020b). The entire data used in
this plot is automatically computed and outputted to the external
\texttt{*\_report.txt} file for each parallel ParaMonte simulation.
\label{fig:procCont512}}
\end{figure}

In (Shahmoradi \& Bagheri, 2020), (Shahmoradi \& Bagheri, 2020b), we
show that the contribution of multiple processors to the construction of
a single Markov Chain in the Fork-Join parallelism paradigm follows a
Geometric distribution. \autoref{fig:procCont512} illustrates an example
contribution distribution of 512 Intel Xeon Phi 7250 processors to the
construction of a single Markov chain exploring a 4-dimensional MVN
target density by the ParaDRAM sampler in the Fork-Join
(\texttt{single-chain}) parallelism mode.

\subsection{Fully-deterministic into-the-future restart
functionality}\label{fully-deterministic-into-the-future-restart-functionality}

A unique feature of the \texttt{ParaMonte::Python} library is the
automatically-enabled fully-deterministic into-the-future restart
functionality of all serial and parallel Monte Carlo simulations. For
example, restarting an interrupted ParaDRAM simulation is as simple as
rerunning the simulation with in its original settings. To ensure a
seamless fully-deterministic restart, all that is required from the user
is to fix the simulation attributes corresponding to the output filename
and the seed of the random number generator of the sampler.

The restart functionality of ParaMonte samplers have been carefully
designed to enable a fully-deterministic reproduction of the originally
interrupted simulation, such that the resulting full chain from the
simulation restart would be the same as the original chain, had the
simulation not been interrupted in the first place. This identity of the
original and restart chains is exact up to 16 digits of decimal
precision.

The restart functionality of \texttt{ParaMonte::Python} samplers is
particularly desired and useful in larges-scale parallel Monte Carlo
simulations and Bayesian inverse problems on supercomputers, given the
fact that the computational time is frequently limited to less than 24
hours for any given simulation. In such cases, all that is needed to
continue a single simulation during multiple separate segments of
allocated supercomputer times, is to simply rerun (restart) the
incomplete simulation.

\section{Documentation and
Repository}\label{documentation-and-repository}

The \texttt{ParaMonte::Python} library presented in this manuscript
builds upon the ParaMonte library for C, C++, and Fortran (Shahmoradi \&
Bagheri, 2020a). Although the Python version of the library has been
only recently released, the kernel samplers of the ParaMonte library
have been already used in number of peer-reviewed publication
(Shahmoradi, 2013), (Shahmoradi, 2013), (Shahmoradi \& Nemiroff, 2014),
(Shahmoradi \& Nemiroff, 2015), (Shahmoradi \& Nemiroff, 2019),
(Osborne, Shahmoradi, \& Nemiroff, 2020), (Osborne, Shahmoradi, \&
Nemiroff, 2020).

An equivalent of \texttt{ParaMonte::Python} is also available for MATLAB
programming environment, with the same set of simulation and
visualization tools and with a syntax and usage that is highly similar
to those of \texttt{ParaMonte::Python}. Consequently, to ensure the
similarity of the ParaMonte package and simulation environment across
different programming languages, we have used and enforced the
\texttt{camelCase} convention in the development of the library in all
programming languages, including Python.

Extensive documentation and examples in Python (as well as C, C++,
Fortran, MATLAB and other programming languages) are available on the
documentation website of the library at:
\url{https://www.cdslab.org/paramonte/}. The ParaMonte library is
MIT-licensed and is permanently located and maintained at
\url{https://github.com/cdslaborg/paramonte/tree/master/src/interface/Python}.

\section{Acknowledgements}\label{acknowledgements}

We acknowledge the use of supercomputing resources at Texas Advanced
Computing Center for the development and testing of
\texttt{ParaMonte::Python} library.

\section*{References}\label{references}
\addcontentsline{toc}{section}{References}

\hypertarget{refs}{}
\hypertarget{ref-Foreman:2019}{}
Foreman-Mackey, D., Farr, W. M., Sinha, M., Archibald, A. M., Hogg, D.
W., Sanders, J. S., Zuntz, J., et al. (2019). Emcee v3: A python
ensemble sampling toolkit for affine-invariant mcmc. \emph{arXiv
preprint arXiv:1911.07688}.

\hypertarget{ref-Haario:2006}{}
Haario, H., Laine, M., Mira, A., \& Saksman, E. (2006). DRAM: Efficient
adaptive mcmc. \emph{Statistics and computing}, \emph{16}(4), 339--354.

\hypertarget{ref-Kumbhare:2020}{}
Kumbhare, S., \& Shahmoradi, A. (2020). Parallel adapative monte carlo
optimization, sampling, and integration in c/c++, fortran, matlab, and
python. \emph{Bulletin of the American Physical Society}.

\hypertarget{ref-Metropolis:1949}{}
Metropolis, N., \& Ulam, S. (1949). The monte carlo method.
\emph{Journal of the American statistical association}, \emph{44}(247),
335--341.

\hypertarget{ref-Miles:2019}{}
Miles, P. R. (2019). Pymcmcstat: A python package for bayesian inference
using delayed rejection adaptive metropolis. \emph{Journal of Open
Source Software}, \emph{4}(38), 1417.

\hypertarget{ref-OsborneARXIV:2020}{}
Osborne, J. A., Shahmoradi, A., \& Nemiroff, R. J. (2020). A multilevel
empirical bayesian approach to estimating the unknown redshifts of 1366
batse catalog long-duration gamma-ray bursts.

\hypertarget{ref-OsborneADS:2020}{}
Osborne, J. A., Shahmoradi, A., \& Nemiroff, R. J. (2020). A Multilevel
Empirical Bayesian Approach to Estimating the Unknown Redshifts of 1366
BATSE Catalog Long-Duration Gamma-Ray Bursts. \emph{arXiv e-prints},
arXiv:2006.01157.

\hypertarget{ref-Patil:2010}{}
Patil, A., Huard, D., \& Fonnesbeck, C. J. (2010). PyMC: Bayesian
stochastic modelling in python. \emph{Journal of statistical software},
\emph{35}(4), 1.

\hypertarget{ref-Shahmoradi:2013}{}
Shahmoradi, A. (2013). A multivariate fit luminosity function and world
model for long gamma-ray bursts. \emph{The Astrophysical Journal},
\emph{766}(2), 111.

\hypertarget{ref-ShahmoradiA:2013}{}
Shahmoradi, A. (2013). Gamma-Ray bursts: Energetics and Prompt
Correlations. \emph{arXiv e-prints}, arXiv:1308.1097.

\hypertarget{ref-ShahmoradiGS:2020}{}
Shahmoradi, A., \& Bagheri, F. (2020). ParaDRAM: A cross-language
toolbox for parallel high-performance delayed-rejection adaptive
metropolis markov chain monte carlo simulations. \emph{arXiv preprint
arXiv:2008.09589}.

\hypertarget{ref-ShahmoradiPMADS:2020}{}
Shahmoradi, A., \& Bagheri, F. (2020a). ParaMonte: A high-performance
serial/parallel Monte Carlo simulation library for C, C++, Fortran.
\emph{arXiv e-prints}, arXiv:2009.14229.

\hypertarget{ref-ShahmoradiADS:2020}{}
Shahmoradi, A., \& Bagheri, F. (2020b). ParaDRAM: A Cross-Language
Toolbox for Parallel High-Performance Delayed-Rejection Adaptive
Metropolis Markov Chain Monte Carlo Simulations. \emph{arXiv e-prints},
arXiv:2008.09589.

\hypertarget{ref-ShahmoradiASCL:2020}{}
Shahmoradi, A., \& Bagheri, F. (2020c, August). ParaMonte: Parallel
Monte Carlo library.

\hypertarget{ref-Shahmoradi:2014}{}
Shahmoradi, A., \& Nemiroff, R. (2014). Classification and energetics of
cosmological gamma-ray bursts. In \emph{American astronomical society
meeting abstracts\# 223} (Vol. 223).

\hypertarget{ref-Shahmoradi:2015}{}
Shahmoradi, A., \& Nemiroff, R. J. (2015). Short versus long gamma-ray
bursts: A comprehensive study of energetics and prompt gamma-ray
correlations. \emph{Monthly Notices of the Royal Astronomical Society},
\emph{451}(1), 126--143.

\hypertarget{ref-Shahmoradi:2019}{}
Shahmoradi, A., \& Nemiroff, R. J. (2019). A Catalog of Redshift
Estimates for 1366 BATSE Long-Duration Gamma-Ray Bursts: Evidence for
Strong Selection Effects on the Phenomenological Prompt Gamma-Ray
Correlations. \emph{arXiv e-prints}, arXiv:1903.06989.

\hypertarget{ref-PyStan:2017}{}
Team, S. D., \& others. (2017). PyStan: The python interface to stan.
\emph{Version 2.16. 0.0}.

\end{document}